# The Hentschel-Procaccia Conjecture

Joseph L. McCauley
Physics Department
University of Houston
Houston, Texas 7204
jmccauley@uh.edu

## Abstract

Extensive confusion exists and persists in the literature on dynamical systems theory, cosmology and other fields over spectra of fractal dimensions. Entirely different generating functions have been treated as if they should yield identical spectra, or scaling exponents. An uncritical implicit assumption of universality of scaling exponents is made, even though there is no good theoretical reason to expect universality away from criticality. Expectations based on the unphysical and empirically inapplicable zero length scale limit (which would require infinite precision in data analysis) are often taken for granted. A source of confusing together different generating functions can be traced back to the Hentschel-Procaccia conjecture, which I prove to be wrong for the case of data analysis in the only applicable limit, that of finite precision. The two different definitions of 'multifractal' stated by Mandelbrot and by Halsey et al are compared and contrasted.

## 1. The empirical distribution and the optimal partition

Mathematical theorems on infinite precision and infinite time limits are generally empirically inapplicable and misleading in the analysis of far from equilibrium dynamical systems. In contrast with our experience with natural phenomena described accurately by equilibrium statistical mechanics, chaotic dynamical systems can generate infinitely many different probability distributions depending on different classes of initial conditions. Theorists are therefore prone to assume that only one invariant measure (the one corresponding to 'measure one initial conditions) is preferred, although there is no known evidence for this in nature. From the standpoint of data analysis the most interesting probability distribution is the empirical distribution P(x), defined by the N data points collected from observation. I will concentrate on the empirical distribution, whatever it is, and for clarity and simplicity of mathematical description I restrict this discussion largely to one dimensional examples.

Consider a (one dimensional) set of data points $\{x_i\}$. The empirical distribution P(x) is simply the fraction of points lying to the left of and including x, so that P(0)=1/N and P(1)=1, by construction. The distribution is constant on the voids and increases discontinuously at each data point, so that the plot of P(x) is a staircase of N-2 steps of finite width. The data staircase has the singular pointwise density $\rho(x) = P'(x)$ given by

$$dP(x) = \frac{dx}{N} \sum_{i=1}^{N} \delta(x - x_i) \quad . \quad (1)$$

Equation (1) already an idealization. It neglects the error bars in the locations of measured positions. In reality each position should be specified empirically by a finite interval whose width is the uncertainty in location, but there is no a priori information on and therefore no precise definition of those error bars. I implicitly assume that the uncertainties in data point locations are very small relative to the smallest separation $l_{min}$ between adjacent data points. Otherwise, coarsegraining and fractal/multifractal analysis are meaningless.

Attempts to 'smooth' the empirical staircase (via splines, e.g.) will discard important information about voids and clustering. *No pointwise probability distribution other than the observed staircase*

$$P(x) = \frac{1}{N} \sum_{j=1}^{N} \theta(x - x_j) \qquad (2)$$

*is relevant for empirical data analysis.*

Contrast this with dynamical systems theory. While probability distributions generated by a chaotic system are not unique, certain chaotic and critical dynamical systems do admit 'generating partitions', which define a hierarchy of unique optimal coarsegrainings of phase space (optimal coarsegrainings of the unique support of all probability distributions generated by the dynamical system). In that case topological universality classes can be defined [1-4] via symbolic dynamics. I call the generating partition the optimal [5] or natural [6] partition because it is the realization of the idea of the infimum in the definition of a multifractafractal point set [7], and because it is generated by the dynamical system. On a generating partition a coarsegrained probability $P_i$ is then defined by the difference $P_i=\Delta P(x)$ over the $i^{th}$ closed interval of size $l_i^{(n)}$, and is just the fraction $P_i=n_i/N$ of the total number of data points $n_i$ in the $i^{th}$ interval including the end points. While a pointwise theoretical distribution $P(x)$ (not unique!) is a staircase of infinitely-many steps, each corresponding coarsegrained distribution $\{P_i\}$ of $P(x)$ is a histogram on the (unique) finite support defined by the hierarchy of intervals defining the generating partition. The generating partition optimally defines the support of the hierarchy of coarsegrained distribution $\{P_i\}$. An arbitrarily constructed partition of empirical data is generally not the generating partition of a deterministic dynamical system and usually cannot yield an accurate estimate of a Hausdorff dimension because of 'convergence problems' arising from the limited number of points in the sample (see part 2 below).

I prefer the frequency definition of probabilities because it arises naturally in both empirical data analysis and computer simulations. An idealized staircase distribution resulting from infinite precision and infinite time limits of a particular dynamical system (the ternary tent map) is provided by the Cantor function [6,8]

$$P(x) = .\frac{\varepsilon_1}{2}\frac{\varepsilon_2}{2}...\frac{\varepsilon_N}{2}... \quad , \quad (3)$$

where, because $x = .\varepsilon_1...\varepsilon_N...$ is a ternary number with $\varepsilon_i = 0$ or 2, $P(x)$ is a binary number (because $\varepsilon_i/2 = 0$ or 1). This staircase describes a mathematician's idealization of empirical data, namely, one distribution (of infinitely-many distributions) consisting of all points in the middle-thirds Cantor set: $P(0)=0$, $P(1)=1$, $P(x)$ is constant on the open voids and increases discontinuously at the end point of any closed interval $l^{(n)} = 3^{-n}$ of the optimal covering, the generating partition of the ternary tent map, where the change in P(x) is in this case $\Delta P = 2^{-n}$. The Cantor function defines a hierarchy of uniform coarsegrained distributions $P_i = \Delta P(x) = N_n^{-1} = 2^{-n}$ on the fractal support $l^{(n)} = 3^{-n}$, for each generation n in the hierarchy. This is by far not the only probability distribution that is generated by the ternary tent map [6]. Multifractal, and also nonuniform, nonfractal distributions are easily generated by the tent map on the same support by other classes of initial data that are easy to find in computation [6].

Summarizing, empirically, I assume only a collection of N raw data points (from a time series, perhaps) generated by a typically unknown deterministic dynamical system, or random process. Only in the former case do we have even the remote possibility of finding a generating partition. We can construct the empirical distribution P(x) immediately from the data, but we cannot construct geometrically/topologically meaningful coarsegrained distributions $\{P_i\}$ unless we can find an optimal partition $\{l_i\}$. In dynamical systems theory an optimal partition is provided by the generating partition, whenever the generating partition exists (tent, logistic, Lozi and Henon maps admit generating partitions, e.g.). The generating partition is the signature of the dynamical system, and shows how the dynamics coarsegrains phase space naturally. The histograms that appear on the optimal support can be produced by every dynamical system in the same topologic universality class: symbolic dynamics is universal for all systems in the same universality class [2-4]. On the other hand, a particular statistical distribution $\{P_i\}$ *cannot* be the signature of a particular deterministic dynamical system [6]. Complex systems generate neither a unique probability distribution nor a generating partition [9,10]. Both the Henon map and the logistic map $f(x) = Dx(1-x)$, with $D_c < D < 4$ where $D_C$ is the period doubling critical point, belong to the same topologic universality class [2] (the logistic map with D>4 and the binary tent map belong to a separate universality class [6]). Both systems, although of different spatial dimension, generate the same range of histograms (for corresponding classes of initial conditions), but on different coarsegrained supports. From the perspective of both dynamical systems theory and the search for scale invariance the central problem of data analysis is to extract the optimal partition from a particular set of data points. See [11,12] for examples of attempts to extract optimal partitions from observational data in fluid mechanics.

## 2. Confusing together different generating functions

The word 'multifractal' is not uniquely defined in the literature. It is sometimes defined, as e.g. in [13], by requiring that the moments of an arbitrary probability distribution P(X,L),

$$\langle X^{q-1} \rangle = \sum_X P(X,L) X^{q-1} \quad , \quad (4)$$

where X is a random value defined in or on intervals of size L, scale like

$$\langle X^{q-1} \rangle \approx L^{\zeta_q} \quad , \quad (5)$$

for small enough interval sizes L (or, in aggregation or cosmology, for large enough L). Without further requirements on X and P(X,L), however, there is no reason to expect that scaling exponents in (5), if they exist at all for a given distribution P(X,L), bear any relation to multifractal spectra of fractal dimensions D($\lambda$) derivable from the generating function

$$Z_n(\beta) = \sum_{i=1}^{N_n} l_i^\beta = \sum_\lambda N(\lambda) l(\lambda)^\beta \approx \sum_\lambda e^{s(\lambda) - \beta\lambda} \quad , \quad (5)$$

which characterizes the fractal via an *optimal partition* {l}, or to generalized dimensions $D_q$ derivable from f($\alpha$) defined by

$$\chi_n(q) = \sum_{i=1}^{N_n} P_i^q \qquad (5b)$$

or

$$\Gamma_n(q) \approx \sum_{i=1}^{N_n} l(\alpha)^{q\alpha - f(\alpha) - \tau} \approx 1 \qquad (5c)$$

where the effects of the optimal partition and coarsegrained probability distribution on that optimal support are tangled together irreversibly. Without further assumptions there is nothing inherently fractal about either X or P(X,L) in (4). More on this below.

The generating function (4) combined with the scaling expectation (5) is used in definitions of multiaffine fractals [14], where a deterministic or random variable (or field) X is continuous but has singular (or no) derivatives. If the distribution of singularities of the field X can be described locally by writing $X \approx L^h$ and $P(X,L) \approx L^{-f(h)}$, then (17) may or may not yield scaling exponents $\zeta_p$ that give rise to a spectrum of generalized dimensions defined by $(p-1)D_p = p\alpha(p) - f(\alpha(p))$ in the infinite precision limit $L \to 0$, where $D_0$, $D_1$, and $D_2$ are fractal *dimensions* of *something* in the model. This holds only when f(h) describes a spectrum of fractal dimensions, but merely calling (4) and (5) 'multifractal' certainly does not make this true. In the latter case, h and f(h) are just a rewriting of a nonfractal probability distribution P(X,L) via a differentiable coordinate transformation, and nonfractal distributions cannot be made fractal (or the converse) by a differentiable coordinate transformation. Stated another way, f(h) is not a spectrum of fractal dimensions of an underlying support of the distribution P unless "L" in (5) represents an optimal or at least efficient partition of the support of the underlying pointwise distribution P(x). Examples in the literature where $X \approx L^h$ with $P(X,L) \approx L^{-f(h)}$ are used without any requirement of the optimal partitioning of a support are height fluctuations in surface roughening [14], self-organized criticality [15] and velocity structure functions in the inertial range of fluid turbulence [16]. For random fractals is no idea of a generating partition, or optimal partition. Examples of multiaffine fractals can be generated both randomly and deterministically [14].

Another source of confusion in the attempt to calculate fractal dimensions and spectra of fractal dimensions, recently in cosmology [17,18] and much earlier dynamical systems theory (before the advent of using the optimal partition [1], especially in the recycling [2] of strange sets, comes from using the correlation integral

$$n(r) = \frac{1}{N}\sum_{i=1}^{N} n_i(r) \quad (6)$$

where

$$n_i(r) = \frac{1}{N}\sum_{\substack{i,j=1 \\ i \neq j}}^{N} \theta(r - |x_i - x_j|) \quad (7)$$

is the fraction of points in a ball of size r, centered on the $i^{th}$ of N points in some distribution. We can always rescale the data to take $0<r<1$. The main point in what follows is that no idea of an optimal partition is required in (6), although (6) is consistent with an optimal partition, if one exists.

In the limit of small length scales r (or, as in cosmology or DLA, in the limit of large r) the generalized dimension $D_2$ coincides with the correlation integral dimension ν whenever

$$n(r) \approx r^\nu \qquad (8)$$

for 'small enough r' whenever an empirical distribution P(x) *exhibits statistical independence* on its *optimal* coarsegrained support [20]. We cannot merely assume statistical independence of observational data. Therefore we can not normally expect to be able to extract $D_2$ from data analysis. Loosely speaking, however, I will still refer to ν as the "correlation dimension" in agreement with widespread usage of that term in the literature, and because empirical data nearly never show statistical independence.

A popular generalization of (6) is defined by the generating function

$$G_n(q) = \frac{1}{N}\sum_{i=1}^{N} n_i(r)^{q-1} \qquad . \qquad (9)$$

As in (6), N is not the number of intervals in a nonoverlapping, efficient (if not optimal) partition but is the number of points in the data sample. *The correlation integral (6) and its generalization (9) have been emphasized in the literature precisely because they appear to allow us to avoid the need to find an optimal partitioning.*

An optimal partition realizes the requirement of imposing an infimum condition [7] on the choice of partition used in calculating the generating functions (5b,c). Avoiding the infimum condition is attractive, because without knowing a generating partition, which can be devilishly hard or impossible to extract from raw data, it's hard to know how to impose the infimum condition. Therefore the popularity of (6) and (9) in data anaylsis. This 'short-circuit', unfortunately, is only an illusion, as I will now explain.

With only very limited data (nearly always the case empirically), the generating function (9) generally cannot be expected to yield either a Hausdorff or box-counting dimension (topologic entropy). Setting q=0 in (9), one might hope that $v_o$ in

$$G_n(0) = \frac{1}{N}\sum_{i=1}^{N} n_i(r)^{-1} \approx r^{-v_o} \qquad (10)$$

would provide an estimate for the box counting dimension $D_o$. This is generally impossible, because neither the box counting nor information dimension is included in the $v_q$ spectrum when there is no underlying generating partition. An appeal to the unphysical limit of vanishing r [21] does not help and is instead very misleading in the empirical case. The reason for the deficit is simple: the terms on the left right side of (10), taken alone without further assumptions, don't define an efficient, nonoverlapping partition of $G_n(0)$ intervals, each of size r. Hence, the "convergence" difficulties reported in [22] in the attempt to estimate the box counting dimension by computing $v_o$ in cosmology, where overlapping intervals occured.

If, following [22], we would try, instead, to define an interval $r_i$ formally by writing $r^v o(n_i(r)^{-1}) = r_i^d$ in (10), then the result

$$\frac{1}{N}\sum_{i=1}^{N} r_i^d \approx 1 \quad (11)$$

reminds us of the definition of the Hausdorff dimension if we would take the infimum over all possible partitions, but in the absence of that condition the exponent d in (11) does not define a Hausdorff dimension: the N intervals $r_i$ may overlap with each other because the sum in [22] is over all N data points instead of over an efficient partition. The problem is that the required infimum condition [7] was not imposed in using the definition (11): Equation (11) was proposed in [23] as one that yields $D_H$, as well as the $D_q$ spectrum for q<1 via the generalization [23b]

$$W_n(t) \approx \frac{1}{N}\sum_{i=1}^{N} r_i^{-t} \approx p^{-q}, \quad (12)$$

but information about the spectrum of generalized dimensions $D_q$, aside from a possible estimate of $D_2$, is not included in these formulae.

Equation (11) as used in [23] is supposed to be based on the equation

$$\frac{1}{N}\sum_{i=1}^{N} r_i \approx \overline{r_N} \quad (13)$$

with

$$\overline{r_N} \approx KN^{-1/D} \quad (14)$$

as used in [24], where $r_i$ is the nearest neighbor distance between two points in a sample consisting of N points. However, in [24] the partitioning used to compute (13) via an example was chosen to be *nonoverlapping*: it is the generating partition of the binary tent map, whereas in [23] an overlapping partition was used implicitly.

In a spirit similar to the attempt to define "multifractal" by using the moments (4) of an arbitrary distribution p(X,L) with a scaling law (5), the definition

$$\langle P^{q-1} \rangle = \sum_P p(P,L) P^{q-1}, \qquad (15)$$

where the probability distribution p(P,L) is arbitrary and is generally undefined, is treated in various places [25] as if it would be identical with the generating function

$$\chi_n(q) = \langle P_i^{q-1} \rangle_{coarsegrained} = \sum_{i=1}^{N_n} P_i P_i^{q-1}, \qquad (5b')$$

although it generally is not due to the lack of an optimal partition in the definition (15).

For an arbitrary probability distribution p(P,L) the two generating functions (15) and (5b') are not even related. Their scaling exponents, if scaling exponents exist at all in either case, are generally not the same, as I explain in the main result in part 3 below. *Multifractal spectra and generalized dimensions are not universal.* Instead, they change with the histograms and their coarsegrained support (partition). I emphasize that it is necessary to think in terms of the *finite precision* limit and to avoid the unphysical limit of vanishing length scale. Otherwise, the results derived may have no empirical applicability. In fractal analysis of observations of nature, everything lies in the finite precision realm. Infinite precision limits combined with appeals to 'finite size effects' ala critical phenomena are inapplicable and misleading in the analysis of far from equilibrium dynamical systems. Resort to thermal equilbrium versions of 'finite size effects' completely irrelevant here.

3. The Hentschel-Procaccia Conjecture

A common source (oft quoted in the literature) of license to confuse together entirely different generating functions can be traced to a speculation based on the zero length scale limit in paper on dynamical systems theory written over sixteen years ago, the Hentschel-Procaccia conjecture. Hentschel and Procaccia boldly suggested [20] (see also [24]) that (5b) is analogous to the Lebesgue integral

$$\langle P^{q-1} \rangle = \int dP(x) P(B_L(x))^{q-1} \qquad (16)$$

and should yield the same generalized dimensions $D_q$ in the small L limit, where P(x) is supposed to be "the natural invariant measure" of a chaotic dynamical system on a strange attractor (see [26] for one definition of "natural measure") and where $P(B_L(x))$ is the fraction of points lying within a ball of size L covering (but not necessarily centered on) a data point x. There are two reasons that may prevent the replacement of (5b) by (16) in data analysis, both centering on lack of uniqueness.

First, there is no empirical evidence that a far from equilibrium dynamical system generates a unique "natural measure" for the various possible initial conditions, meaning "present conditions" found in nature. *Mathematically* seen, an isolated chaotic dynamical system can generate infinitely many different distributions (probability measures) P(x) for infinitely many different classes of initial conditions [5,6]. Empirically, this dependence of observed dynamics on initial conditions does not create a difficulty: empirical data are described by the staircase function

$$P(x) = \frac{1}{N} \sum_{j=1}^{N} \theta(x - x_j) \qquad (2)$$

where the $\{x_i\}$ are the observed data points. Without making any special theoretical assumptions that might bias a data analysis, we can simply say that the initial conditions, *whatever* they were, produced the empirical distribution P(x) via the time evolution of the dynamical system, whatever it is. Furthermore, the details of the empirical distribution (2) could even be eliminated by using the generating function (5) to extract D(λ) from the data, if a generating partition can be found from the data. Now for my main point.

Given an empirical measure (3b), and even a possible underlying explanatory dynamical model with a generating partition, there is *still* ambiguity inherent in the attempt to use (16) as a replacement for (5b). In finite precision there are infinitely many different possible results the integral (16), depending on *which* subset of the data set we decide to measure. In particular, before we can identify the function $P(B_L(x))$ we must first define what we mean by "L". Here, I consider the two main possibilities discussed in the literature. These two possibilities were conjectured [20] (before the discovery of 'recycling' [2]) to yield the same result in the infinite precision limit (where L vanishes), but I will now show that they do not yield the same result in the finite precision regime. Please bear in mind that I have explained above why, when "L" is finite, (5b) and (9) yield entirely different spectra of scaling exponents, if scaling exists at all.

If we choose the balls/intervals $B_L(x)$ to have arbitrary length L centered on a data point $x_i$ (as in [27]), then the fraction of points lying within each interval of size L is given by $P(B_L(x)) = n(x,L)$ where $n(x_i,L) = n_i(L)$ with

$$n_i(L) = \frac{1}{N} \sum_{\substack{i,j=1 \\ i \neq j}}^{N} \theta(l - |x_i - x_j|)$$

. (7b)

If we use the pointwise definitions $P(B_L(x)) = n(x,L)$ and

$$dP(x) = \frac{dx}{N} \sum_{i=1}^{N} \delta(x - x_i) \quad (1)$$

in the integral (16) then we obtain the correlation integral generating function

$$\int dP(x) P(B_l(x))^{q-1} = \frac{1}{N} \sum_{i=1}^{N} n_i(l)^{q-1} = G_n(q)$$

, (9b)

which differs significantly from (5b) in data analysis because it makes no reference to an optimal partition and therefore cannot be used to find Hausdorff (or even box counting) dimension for arbitrary (non-optimal partionable) data.

As an aside, in dynamical systems theory the N intervals can in principle be chosen small enough not to overlap with each other: ideally, on a mathematically-defined strange attractor there exist in principle $t^\infty$ points in any neighborhood of any arbitrary point $x_i$ on the attractor, where $t^\infty$ is the cardinality of the attractor. Here, as opposed to applications to cosmology, the N intervals (or balls) $B_L$ of size L can be chosen small enough not to overlap, but generally do not partition the attractor optimally. The topologic entropy can be calculated from N and t. Equation (6), which was not invented with partitioning in mind, defines a time average over N points on the attractor, and the uniform weight 1/N is correct because each point $x_i$ occurs exactly once (so long as trajectories of the dynamical system are unique, which we assume here). In calculations the number N of points can be increased by increasing the precision of the calculation. In cosmology, in contrast, N is the total number of galaxies in a finite sample, so that the N intervals of size l are always overlapping, at least with currently-available data [28].

The important point is that there is a different way to define balls $B_L(x)$ and a corresponding measure $P(B_L(x))$. *The choice of which function to integrate with respect to the measure is not unique*. Instead of choosing N uniform intervals where N is the number of data points, requiring the pointwise definition $P(B_L(x)) = n(x,L)$ as given by (7b) (which does *not* include or depend on a partition of the data set), we can instead choose our balls $B_L$ to be the $N_n$ intervals $\{l_i\}$ in the optimal partition of the data, *if* the data are generated by a dynamical system with a generating partition. In this case $P(B_L(x))$ is given by the simple function

$$P(B_l(x)) = \sum_{i=1}^{N_n} P_i \chi_{l_i}(x) \qquad (17)$$

where $\chi_{l_i}(x)$ is the characteristic function for the partition $\{l_i\}$ of disjoint intervals [29] and

$$P_i = P(x_i + l_i) - P(x_i) = \frac{1}{N} \sum_{j=i}^{i+n_i-1} \theta(x_{i+n_i} - x_j) = \frac{n_i}{N} . \qquad (18)$$

Here, $x_i$ and $x_{i+n_i} = x_i+l_i$ are taken to be the end points of any of the $N_n$ *optimal* intervals $\{l_i\}$. With this optimal choice of "what to measure" (optimal choice of function $P(B_L(x))$ to integrate with respect to the measure $P(x)$) the integral (16) yields

$$\int dP(x) P(B_l(x))^{q-1} = \sum_{i=1}^{N_n} P_i P_i^{q-1} = \chi_n(q) . \qquad (5b'')$$

*From the standpoint of both data analysis and measure theory the only significant difference between the distributions (5b) and (9) is the lack of a partition in (9), and the use of an optimal partition to define (5b). Whether these two generating functions do or do not, in the limit of $l^{(n)} \to 0$ for a mathematical fractal of cardinality $t\infty$, yield the same generalized dimensions (whether $D_q = v_q$ as $l^{(n)} \to 0$) is completely irrelevant for the correct analysis of empirical data.*

### 4. Two main definitions of 'multifractal'

A Cramer function [16,30] provides a systematic way of obtaining a universal result $X \approx L^h$ and $P(X,L) \approx L^{-f(h)}$ via a limit theorem in classical statistics, *for the case where the $h_i$ exist and are independent random variables.* This result has no necessary connection with the idea of spectra of fractal dimensions $D(\lambda)$, or generalized dimensions $D_q$ derivable from spectra of fractal dimensions $f(\alpha)$. The Cramer function, emphasized by Mandelbrot and Frisch, is based on the law of large numbers and may appear in completely nonfractal contexts, such as classical statistical mechanics. The generating functions (5b) and (5c) were, in contrast, invented to handle singularities of probability distributions, especially as appear in deterministic dynamics on a Cantor set (the condition of the infimum [7] required in defining (5b,c) is satisfied by a generating partition). However, in a paper that includes one of the coauthors of Halsey et al [30] as author, the Cramer function was used, was called "multifractal", and then the paper by Halsey et al [30] was cited as the reference [15]. However, the Cramer function is not defined or mentioned in [30]!

One can choose to follow Halsey et al [30] in defining $f(\alpha)$ and $D_q$, or one can follow Mandelbrot [29] in defining "$v(\alpha)$ and $d_q$" via Cramer functions [16], but one should not mix the two separate definitions together without comment as is done in [13,15]. I recommend the definition of multifractal given in [30] as the standard because, in that case, $f(\alpha)$ is *always* the Hausdorff dimension of a subset of the support of a distribution $P(x)$. The necessity of an optimal partition in order to define $f(\alpha)$ was stated in Halsey et al [30] (see their "infimum" requirement), but was not emphasized strongly enough within the fractals community at that time. The role played by the generating partition in defining fractal attractors became clear only after later advanced work in dynamical systems theory [1,2,4]. Mandelbrot originally proposed the Cramer function as a replacement for $f(\alpha)$ because the infimum condition required to calculate Hausdorff dimensions [7] is so hard to implement in practice, and also because he so often emphasizes randomness (as opposed to determinism) in his discussions of fractals (in the finance, e.g., where evidence for randomness abounds while evidence for determinism has not yet been found).

Spectra of fractal dimensions that satisfy the infimum condition have, nevertheless been calculated via generating partitions [1,4]. Empirical data have been analyzed on the same basis [11]. The large n limit (or small L limit) may no easier to implement in practice than is the problem of finding an efficient partition. In many cases evidence for neither can be found, but the 'convergence' difficulties that generally occur when trying to compute fractal dimensions via box-counting by using an arbitrary partition can only be reduced or eliminated by the use of the optimal or at least efficient [28] partition.

**Acknowledgement**

This work was performed during my 1986-87 sabbatical at Ludwig-Maxmillian-Universität (München) where, with strong encouragement of from Herbert Wagner, I had the good luck to collaborate with Martin Kerscher on the analysis of galaxy statistics. I am grateful to Wagner for very fine hospitality, including the occupation of a free Lehrstuhl office, and to both Wagner and Kerscher for the stimulating discussions that led to this work. Without Kerscher's gentle prodding, I would not have opened Falconer's book, and would not have learned that abstract probability measure theory is general enough to include the case of finitely many empirical data points. That was the key to the analysis of the Hentschel-Procaccia conjecture.